\documentclass[aoas,nameyear,dvips]{arximspdf}
\usepackage{dcolumn}
\usepackage{graphicx}

\doi{10.1214/09-AOAS268}
\volume{4}
\issue{1}
\pubyear{2010}
\firstpage{26}
\lastpage{52}

\makeatletter
\newcolumntype{t}[1]{D{,}{}{#1}}
\newcolumntype{d}[1]{D{.}{.}{#1}}
\makeatother

\begin{document}
\begin{frontmatter}

\title{Analysis of dependence among size, rate and duration in internet flows}
\runtitle{Dependence in internet flows}

\begin{aug}
\author[A]{\fnms{Cheolwoo} \snm{Park}\ead[label=e1]{cpark@stat.uga.edu}\corref{}},
\author[C]{\fnms{Felix} \snm{Hern\'{a}ndez-Campos}\thanksref{t1}\ead[label=e2]{fhernand@cs.unc.edu}},
\author[B]{\fnms{J. S.} \snm{Marron}\ead[label=e3]{marron@email.unc.edu}},\\
\author[C]{\fnms{Kevin} \snm{Jeffay}\thanksref{t2}\ead[label=e4]{jeffay@cs.unc.edu}}
\and
\author[C]{\fnms{F. Donelson} \snm{Smith}\thanksref{t2}\ead[label=e5]{smithfd@cs.unc.edu}}
\thankstext{t1}{Currently at Google, Inc.}
\thankstext{t2}{Supported in part by NSF CRI 07-09081.}
\runauthor{C. Park et al.}
\affiliation{University of Georgia, University of North Carolina at
Chapel Hill,
University of North Carolina at Chapel Hill,
University of North Carolina at Chapel Hill and University of North
Carolina at Chapel Hill}
\address[A]{C. Park\\
Department of Statistics\\
University of Georgia\\
Athens, Georgia 30602\\ USA\\
\printead{e1}}
\address[B]{J. S. Marron\\
Department of Statistics\\
 \quad and Operations Research\\
University of North Carolina\\
 \quad at Chapel Hill\\
Chapel Hill, North Carolina 27599\\ USA\\
\printead{e3}}
\address[C]{F. Hern\'{a}ndez-Campos\\
K. Jeffay\\
F. D. Smith\\
Department of Computer Science\\
University of North Carolina\\
 \quad at Chapel Hill\\
Chapel Hill, North Carolina 27599\\ USA\\
\printead{e2}\\
\phantom{E-mail: }\printead*{e4}\\
\phantom{E-mail: }\printead*{e5}}
\end{aug}

\pdfauthor{Cheolwoo Park, Felix Hernandez-Campos, J. S. Marron, Kevin Jeffay, F. Donelson Smith}

\received{\smonth{12} \syear{2008}}
\revised{\smonth{6} \syear{2009}}

%
\begin{abstract}
In this paper we examine rigorously the evidence for dependence
among data size, transfer rate and duration in Internet flows. We
emphasize two statistical approaches for studying dependence,
including Pearson's correlation coefficient and the extremal
dependence analysis method. We apply these methods to large
data sets of packet traces from three networks. Our major
results show that Pearson's correlation coefficients between size
and duration are much smaller than one might expect. We also find
that correlation coefficients between size and rate are generally
small and can be strongly affected by applying thresholds to size or
duration. Based on Transmission Control Protocol connection startup
mechanisms, we argue that thresholds on size should be more useful
than thresholds on duration in the analysis of correlations. Using
extremal dependence analysis, we draw a similar conclusion, finding
remarkable independence for extremal values of size and rate.
\end{abstract}

%

\end{frontmatter}
%
\section{Introduction and background}
\label{sec:intro}

In today's Internet approximately 90\% of all data is transmitted between
applications that rely on Transmission Control Protocol (TCP) connections
to provide reliable, in-order delivery of that data. A number of
important issues have been raised
concerning the existence of correlations between the amount of data
transferred over
a TCP connection (its size) and the duration of the connection or the
rate at which the data is transferred.
Intuitively, we would expect there to be a strong positive correlation
between the size
of data and the transfer's duration since larger amounts of data should
require more time to transfer at any given transfer rate.
Similarly, there should be a strong negative correlation between the
transfer rate and the duration
since faster transfer rates should result in shorter connection
durations for any given data size.
In this paper we revisit the issue of dependence among sizes, durations
and rates.

It is the presence (or absence) of correlation between the amount of
data transferred and the rate at
which it is transferred that has the most important implications for
networking operations.
We might find that the amount of data transferred is largely
independent of the rate at which it is transferred.
This could happen because the amount of data transferred in a TCP
connection is the result of an explicit request
by a user for some data object of a certain size and having some
intrinsic utility for that user.
The transfer rate, however, is determined by independent network
factors such as the bottleneck link bandwidth,
the maximum window size, the round-trip time and the loss rate.

Suppose instead we find that size and rate are strongly correlated.
What might the implications be? One obvious possibility is that this
correlation was caused primarily
by users' behaviors in choosing the amount of data they transfer based
on their expectations of the rate
at which it would be transferred. To quote directly from one study that
found empirical evidence of
a strong positive correlation between size and rate: ``This is strong
evidence that user behavior,
as evidenced by the amount of data they transfer, is not intrinsically
determined, but rather,
is a~function of the speed at which files can be downloaded'' [\citet{r23}, page~309].
``Thus, users appear to choose the size of their transfer based,
strongly, on the available bandwidth''
[\citet{r23}, page 313].

If it can be shown that network users do, in fact, mostly select data
objects based on expectations of the data transfer rate,
there may be significant implications for future network growth and
provisioning.
If faster network technologies are deployed at the network edge (e.g.,
fiber to the home),
users may choose to download larger data objects than when using slower
edge links (cable modems, DSL).
This in turn opens the possibility of new applications and services
that inherently require larger data objects
(e.g., videos in web pages, HD-DVD downloads). Conversely, if high
bandwidth is not available to users,
they may choose to forgo accessing these large objects since their size
implies longer durations,
and that may outweigh other intrinsic values. Large investments in
network-edge technologies
and backbone capacity by Internet service providers (ISPs) and content
providers may be influenced by their perceptions of
how strongly users react to their experiences that relate the intrinsic
value of large data objects
to the rate at which they are transferred.

Unfortunately, analyzing correlations
in real-world TCP connections is difficult. The few published studies
have produced mixed and
sometimes conflicting results; these studies are reviewed briefly in
Section \ref{sec:lit}.
The analysis of TCP correlations is complicated by several TCP design
factors, so a
gentle introduction to them is given in Section~\ref{sec:data}.
In this paper we use different data sets described in Section~\ref{sec:data}
consisting of TCP traces from three networks: a backbone link in the
Internet/2 core (the Abilene network),
and two access links---one from Bell Labs and another from the
University of North Carolina at Chapel Hill (UNC).

In our examination of these data sets, we used both Pearson's
correlation coefficient
and a two-dimensional analysis of joint thresholds on both size and
duration on a log scale.
We used Pearson's correlation because we wanted to compare our results
with those of \citet{r23},
which motivated this study, and drew a different conclusion using this
simple method.
In Section \ref{sec:corrt} we examined both the collection of all TCP
connections in the traces and the subset
consisting of only those used for web browsing. The latter subset
represents a particular type of
network application where users make choices of which data content to transfer
and, thus, we might expect to see more correlation of size and rate.
In correlation analysis, the choices of threshold to segment the data
have influences on the results and
this issue is addressed using both real network traces and simulated
data. The simulated data were generated
from a bivariate normal distribution, which was different from the
actual distribution of
size and duration (or rate) on a~log scale. However, the thresholding
effect was still valid for the simulated data,
which implies that it is not data specific.

Many application protocols, notably the HyperText Transfer Protocol
(HTTP) used by web browsers and servers,
reuse an already established TCP connection (called a persistent
connection) to transfer many data
objects of various sizes. Each of these data objects might be chosen
individually
by the user, while the total duration and number of objects transferred
on a persistent connection is
more typically the result of constraints on resources such as memory at
the server or browser program.
To understand the potential differences in correlation between rates
and sizes of connections vs.
rates and sizes of data objects, we extracted the sizes of individual
data objects transferred and
repeated the analysis. These results are presented in Section \ref{sec:corrd}.
Finally, in Section \ref{sec:eda}, we applied an extremal dependence
analysis method proposed
by \citet{r8} to study the dependence among size, duration and rate.
Section \ref{sec:summary}
gives conclusions based on our results and discusses their implications.

\section{Related work}
\label{sec:lit}

The first large-scale investigation of correlations among size, rate
and duration in TCP was done by \citet{r23}.
The data they analyzed came from packet traces from a mixture of
network access and backbone links.
The total data set represented over 20 billion packets. Their results
are expressed in terms of correlations within TCP flows.
In their analysis, they defined one flow for each unidirectional
transmission of packets between the endpoints of a connection.
In a very large number of network application protocols, data objects
are transferred in only one direction of flow
in the connection, while only small packets containing acknowledgments
and control information are transferred
in the reverse flow direction. Because of this issue, we focused in
this study on TCP connections (bidirectional transfers of data between
endpoints in an established TCP connection).
We use connections instead of flows because they correspond more
closely to the user and application
concepts of complete data transfers.

Correlations among the logarithms of flow size, rate and duration were
computed because of the wide range
and skewed distributions of these metrics. Thresholds on durations were
used to segment the data.
For flows with durations longer than 5 seconds, they found the
following results:
\begin{enumerate}[]
\item[$\bullet$] Slight negative correlation between duration and rate (range
across data sets of $-0.187$ to $-0.453$),
\item[$\bullet$] Slight positive correlation between size and duration (0.10 to
0.296), and
\item[$\bullet$] Strong positive correlation between size and rate (0.835 to 0.885).
\end{enumerate}

The strong correlation of size and rate was also found when other
values of the duration threshold were used.
For flows with durations longer than 1 second, the correlations ranged
from 0.65 to 0.77 and for those longer than
30 seconds, the range was 0.90 to 0.95. This increasing correlation
strength with increasing duration threshold
appeared to reinforce their case for users' behavior that was dependent
on expectations about available network transfer rates.

\citet{r13} also investigated the correlations among size, rate and
duration in Internet flows using packet traces
from regional and national backbone networks. They used the same
definitions of flows (two unidirectional flows
for each TCP connection), size, duration and rate as Zhang, Breslau and Shenker~(\citeyear{r23}) and
computed correlations using Kendall's
$\tau$ method (which they claim is more robust to outliers and
non-normality than Pearson's method).
Without using any threshold on duration or when using the same
thresholds on duration (1 second, 5 seconds and 30 seconds)
as \citet{r23}, they also found that size and rate were strongly
correlated. Interestingly, when they used size
as a threshold to segment the data, they found that size and rate
correlations essentially disappeared or became very weak.
They also found that most of the flows with durations longer than 30
seconds were actually relatively small in size
(70\% were smaller than 10 kB). By examining many of the flows in
detail, they concluded that even flows
with relatively long durations can be sufficiently small in size so
that TCP protocol mechanisms are the primary source
of correlations. Further, they found that the correlation of size and
rate was strongly influenced
by the TCP mechanisms even with sizes above 10 kB.

A third study that considered the correlation of size with rate and
duration is presented by \citet{r16}.
In this study, active measurements were done using the PlanetLab
network to transfer various sizes of data
using TCP connections and record the transfer times. As in the prior
two studies, they found a strong correlation
between size and rate (0.80 to 0.95) for file transfers using sizes
that ranged from 5 kB to 1 GB.
Their analysis concluded that most of the correlation was due to
startup overheads and ``residual'' effects of TCP protocol mechanisms.

Taken together, the results from these three studies raise the
possibility that a~strong correlation exists
between size and rate but not between size and duration in TCP
connections. The results are, however,
not completely consistent among studies. Further, it is not clear
whether the correlations are best explained
by TCP protocol effects or as the result of user choices and actions.

\section{Data sets and methods}
\label{sec:data}

The one-hour Abilene-I trace was collected at a~2.5 Gbps link between
Indianapolis and Cleveland.
The capture took place on August 14, 2002, between 9 AM and 10 AM. It
represents TCP connections among hundreds of universities
and research institutions using a very heterogeneous set of
technologies at the network's edge. The link that was traced for the
Bell Labs data
connected the Murray Hill facility (a population of about 450 technical
and administrative staff) to the Internet. The trace captures 168 hours
of operation
between May 19 and May 25, 2002. While the traced link was a 100 Mbps
Ethernet carrying the Lab's aggregated
traffic to its router, the router's outbound link to the Internet was
limited to 9 Mbps.

The third trace was a one-hour trace of the 1 Gbps Ethernet link
connecting the campus of the University of North Carolina
at Chapel Hill (UNC) with its Internet router (this router's outbound
link to the Internet operated at 2.5 Gbps speed).
The capture took place on April 30, 2003 at 7 PM.
A population of over 40,000 users including students, faculty, and
administrators from
academic departments, research institutions, and a medical complex
used this link for Internet connectivity.
These three traces are contemporary with \citet{r23}. While they are
somewhat old, the primary features of
TCP that control transfer rate (i.e., slow start and congestion
avoidance) remain basically unchanged. Furthermore, the bandwidths
available to today's home and wireless users are rapidly approaching
those available to the users of the networks considered in our study
(i.e., 10 to 100 Mbps).

The summary statistics for these three traces are given
in Table \ref{tab:summary}. We identified the subset of
our TCP connections that are likely to have been used for the web
protocol HTTP. The summary statistics for the TCP
and HTTP connections are also given in Table \ref{tab:summary}.
As Table \ref{tab:summary} clearly shows, we are using large samples
(over one million in every case)
of TCP and HTTP connections in this analysis. Further, Table \ref
{tab:summary} shows that these TCP connections
are transferring large amounts of data including substantial
proportions of web (HTTP) traffic.

\begin{table}[b]
\caption{Summary statistics for the traces}
\label{tab:summary}
\begin{tabular*}{\textwidth}{@{\extracolsep{\fill}}lccccc@{}}
\hline
& &  \multicolumn{2}{c}{\textbf{TCP}} & \multicolumn{2}{c@{}}{\textbf{HTTP}}\\[-6pt]
& &  \multicolumn{2}{c}{\hrulefill} & \multicolumn{2}{c@{}}{\hrulefill}\\
\textbf{Trace} & \textbf{Packets} & \textbf{Connections} & \textbf{Bytes} & \textbf{Connections} & \textbf{Bytes} \\
\hline
Abilene & 887.47 Million & 1,318,661 & 334.27 GB & 1,003,817 & 69.38 GB
\\
Bell Labs & \phantom{0}27.93 Million & 2,313,744 & \phantom{0}80.80 GB & 1,967,442 & 27.26
GB\\
UNC & 109.80 Million & 1,433,924 & 120.35 GB & 1,055,823 & 37.33 GB\\
\hline
\end{tabular*}
\end{table}

We applied a number of trace processing tools used in \citet{r7} to
extract the necessary data on sizes, durations and rates.
We eliminated connections with 0 second durations (a single packet).
For each TCP connection we define its duration
as the difference between the timestamps of the first and the last
packet seen for this connection in the trace.
The size of a TCP connection is the total number of bytes transferred
by this connection (in both directions).
The log (base 10) of both duration and size are computed for each TCP
connection and used in the analysis.

Next we provide a brief, high-level introduction to the mechanisms in
the TCP protocols [see \citet{r12}]
that may introduce correlations among size, rate and durations.
Our purpose in this discussion is to clarify why it may be necessary to
use certain thresholds in the analysis.

TCP sends data using a dynamically changing congestion window $W$,
which limits the maximum amount of unacknowledged data
in the network to $W$ packets. Since acknowledgments from the receiver
take a minimum of one round-trip time, $T$,
to reach the sender of the data, TCP cannot send data faster than
$W/T$. When the connection is established,
the size of $W$ is first set to one packet (typically 1460 bytes of
data), and it is doubled after each window of packets
is fully acknowledged. This means that the rate of a TCP connection
during startup is correlated with the size of the data transferred,
since the more data there is to transfer, the greater the opportunity
TCP has to further increase $W$ and therefore the rate.
If a TCP connection is used to transfer an amount of data such that it
continues to double $W$ every time interval,
$T$, then there are correlations among size, rate and duration. This,
however, is correlation caused purely by TCP behavior
and does not reflect any influence by users' choices of data objects
based on experienced transfer rates.

The window-doubling mechanism is used until either (a) there is a
packet loss, (b) $W$ reaches a configured maximum at the sender
(up to 1 MB), or (c) $W$ reaches the size of the receiver's current
buffer space available to hold packets (64 kB or less).
Under normal (no loss) conditions, TCP transitions to increasing $W$ by~1 packet for each fully acknowledged window of packets
after the threshold described in (b) above is reached. However, TCP
cannot send at a rate that would cause the receiver's
buffer space to overflow, so the amount of unacknowledged data the
sender can send into the network is constrained to be MIN
($W$, receiver's buffer space). If $W$ becomes larger than the
receiver's buffer space, the actual sending rate of the
connection cannot increase and, therefore, the rate becomes stable and
independent of the size of the data transferred.
We have considered so far the rate of data units sent from the start of
a TCP connection.
The dependency between size and rate changes when the start of the data
transfer is some point in the middle of a connection
(e.g., the second and later data transfers in a persistent web browser
connection). In this case,
$W$ can be much larger at the start of the transfer than in the case of
a transfer at the beginning of the TCP connection.

We cannot use measurements of rate and sizes near the beginning of a
connection, since TCP itself, without any user interaction,
creates a strong correlation between size and rate. This TCP induced
correlation tends to not be present in longer transfers
because of the limiting effect of relatively small receiver buffer space.
Thus, analyzing measurements of larger data transfers is the correct
way to study possible correlations with data size.
Thresholds on duration alone are not sufficient. Based on TCP
connection startup mechanisms, we argue that thresholds
on size should be more useful than thresholds on duration in the
analysis of correlations.

\section{Correlation analysis of TCP connections}
\label{sec:corrt}

In this section we examine correlations of size, rate and duration
considering the total size of data transferred on
individual TCP connections. We compute Pearson's correlation
coefficients of the log of these three variables
as in \citet{r23} because of the large range and heavy tailed distribution.
We first consider the correlation of size and duration where the
definitions of those metrics
are those given in Section~\ref{sec:data}. One observation from the
prior work reviewed in Section~\ref{sec:lit}
is that the choice of threshold to segment the data has influences on
the results. We address this issue by performing
an analysis of thresholds in two dimensions (size and duration)
considered jointly. The main idea is to embed thresholds
used in prior work in a larger, more complete analysis where we can
study the effect as thresholds are varied in two dimensions.

\begin{figure*}
\centering
\begin{tabular}{c}

\includegraphics{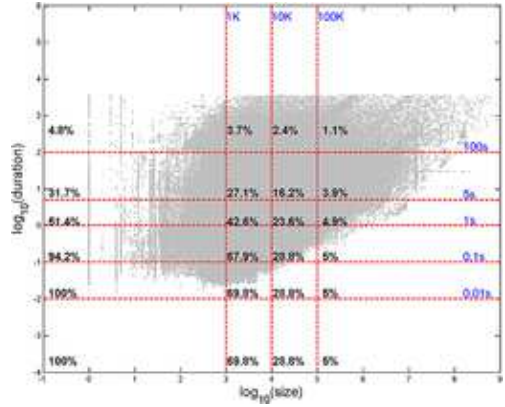}
\\
\footnotesize{(a) All Abilene connections, $n=1\mbox{,}318\mbox{,}661$.}\\[3pt]

\includegraphics{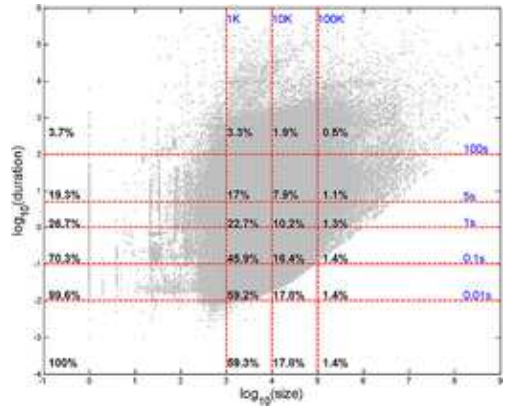}
\\
\footnotesize{(b) All Bell Labs connections, $n=2\mbox{,}313\mbox{,}744$.}\\[3pt]

\includegraphics{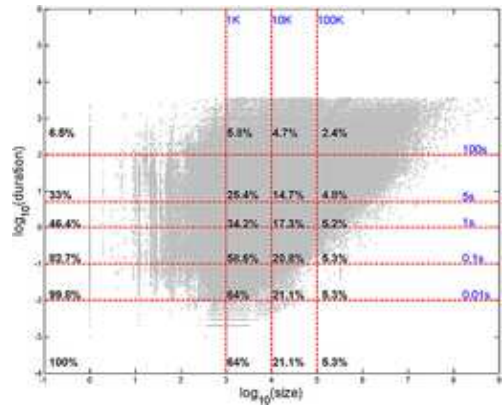}
\\
\footnotesize{(c) All UNC connections, $n=1\mbox{,}433\mbox{,}924$.}
\end{tabular}
\caption{Duration--size log--log correlation.}\label{fig:sd}
\end{figure*}

Figure \ref{fig:sd} gives scatter plots of $\log_{10}$(duration) vs.
$\log_{10}$(size) where each point represents one TCP connection
found in the respective trace. Also shown on the plots are
horizontal lines (showing how the population of TCP connections is
divided into subpopulations by duration thresholds) and vertical
lines (showing how the population is divided by size thresholds).
All points above a horizontal line represent TCP connections with
durations greater than the indicated threshold. Note that in
Figures \ref{fig:sd}(a) and (c) no $\log_{10}$(duration) value is
greater than 3.5 because these two traces are only 3600 seconds in
duration. The Bell Labs trace was more than 2 orders of magnitude
longer so $\log_{10}$(duration) values $> 5$ were observed.

All points to the right of a vertical line represent TCP connections
with total size greater than the indicated threshold.
The intersections of the threshold lines with each other or with the
horizontal (bottom) and vertical (left)
axes define subpopulations that include all the points above (longer
duration) and to the right (larger size)
of the intersection point. Near each such intersection is a number that
gives the percentage of the total population
that falls in the subpopulation.
%
\begin{table}
\caption{Log--log correlation coefficients of size and duration for all
TCP connections}\label{tab:corr1}
\begin{tabular*}{\textwidth}{@{\extracolsep{\fill}}lt{1.5}cccc@{}}
\hline
& \multicolumn{1}{c}{\textbf{Duration}\hspace*{3.9pt}} & \textbf{Size} & \textbf{Size} & \textbf{Size} & \textbf{Size} \\
& \multicolumn{1}{c}{\textbf{(seconds)}\hspace*{3.9pt}} & $\bolds >$\textbf{0 kB} & $\bolds >$\textbf{1 kB} & $\bolds >$\textbf{10 kB} & $\bolds >$\textbf{100 kB}\\
\hline
Abilene & >,100.0 & 0.081 & 0.183 & 0.204 & 0.121 \\
& >,5.0 & 0.010 & 0.225 & 0.281 & 0.254 \\
& >,1.0 & 0.180 & 0.253 & 0.301 & 0.295 \\
& >,0.10 & 0.415 & 0.445 & 0.376 & 0.310 \\
& >,0.01 & 0.453 & 0.465 & 0.376 & 0.310 \\
& >,0.0 & 0.453 & 0.465 & 0.376 & 0.310 \\[5pt]
Bell Labs & >,100.0 & 0.070 & 0.030 & 0.106 & 0.128 \\
& >,5.0 & 0.159 & 0.190 & 0.214 & 0.139 \\
& >,1.0 & 0.210 & 0.177 & 0.177 & 0.134 \\
& >,0.10 & 0.396 & 0.324 & 0.235 & 0.198 \\
& >,0.01 & 0.434 & 0.432 & 0.278 & 0.198 \\
& >,0.0 & 0.436 & 0.433 & 0.278 & 0.198 \\[5pt]
UNC & >,100.0 & 0.204 & 0.145 & 0.287 & 0.305 \\
& >,5.0 & 0.351 & 0.420 & 0.408 & 0.386 \\
& >,1.0 & 0.361 & 0.459 & 0.408 & 0.411 \\
& >,0.10 & 0.394 & 0.552 & 0.471 & 0.413 \\
& >,0.01 & 0.451 & 0.572 & 0.480 & 0.413 \\
& >,0.0 & 0.452 & 0.572 & 0.480 & 0.413 \\
\hline
\end{tabular*}
\end{table}

The size-duration correlation (computed as the Pearson's correlation
coefficient on a log scale) for each of the subpopulations
shown in Figure \ref{fig:sd} is given in Table~\ref{tab:corr1}.
Considering all connections (duration $>0$ and size $>0$),
we find only a weak positive correlation between size and duration that
is remarkably consistent across traces
(0.453, 0.436 and 0.452). When using only the 5-second duration
threshold from \citet{r23}, we find a wider range
(0.10, 0.159 and 0.351) but generally weaker correlation.

\begin{table}
\caption{Log--log correlation coefficients of size and duration for all
HTTP connections}
\label{tab:corr2}
\begin{tabular*}{\textwidth}{@{\extracolsep{\fill}}lt{1.5}d{3.4}d{3.4}d{3.4}d{3.4}@{}}
\hline
& \multicolumn{1}{c}{\textbf{Duration}\hspace*{3.9pt}} & \multicolumn{1}{c}{\textbf{Size}} & \multicolumn{1}{c}{\textbf{Size}}
& \multicolumn{1}{c}{\textbf{Size}} & \multicolumn{1}{c@{}}{\textbf{Size}} \\
& \multicolumn{1}{c}{\textbf{(seconds)}\hspace*{3.9pt}} & \multicolumn{1}{c}{$\bolds >$\textbf{0 kB}}
& \multicolumn{1}{c}{$\bolds >$\textbf{1 kB}} & \multicolumn{1}{c}{$\bolds >$\textbf{10 kB}}
& \multicolumn{1}{c@{}}{$\bolds >$\textbf{100 kB}}\\
\hline
Abilene &>,100.0 & -0.269 & 0.056 & 0.089 & 0.188 \\
& & (3.6\%) & (3.2\%) & (2.0\%) & (0.6\%)\\
& >,5.0 & 0.029 & 0.123 & 0.188 & 0.191 \\
& & (32.3\%) & (29.3\%) & (18.2\%) & (3.4\%)\\
& >,1.0 & 0.111 & 0.165 & 0.255 & 0.246 \\
& & (48.2\%) & (43.3\%) & (26.3\%) & (4.1\%)\\
& >,0.10 & 0.474 & 0.460 & 0.354 & 0.256 \\
& & (94.2\%) & (72.1\%) & (32.2\%) & (4.1\%)\\
& >,0.01 & 0.506 & 0.486 & 0.354 & 0.256 \\
& & (100.0\%) & (74.5\%) & (32.2\%) & (4.1\%)\\
& >,0.0 & 0.506 & 0.486 & 0.354 & 0.256 \\
& & (100.0\%) & (74.5\%) & (32.2\%) & (4.1\%)\\[5pt]
Bell Labs &>,100.0 & 0.089 & 0.038 & 0.048 & -0.021 \\
& & (3.7\%) & (3.4\%) & (2.0\%) & (0.4\%)\\
& >,5.0 & 0.148 & 0.142 & 0.176 & 0.052 \\
& & (17.6\%) & (15.3\%) & (8.0\%) & (1.0\%)\\
& >,1.0 & 0.171 & 0.127 & 0.142 & 0.071 \\
& & (23.6\%) & (20.0\%) & (10.3\%) & (1.1\%)\\
& >,0.10 & 0.399 & 0.344 & 0.222 & 0.144 \\
& & (65.7\%) & (46.3\%) & (17.2\%) & (1.2\%)\\
& >,0.01 & 0.492 & 0.454 & 0.269 & 0.144 \\
& & (99.6\%) & (61.6\%) & (18.8\%) & (1.2\%)\\
&>,0.0 & 0.494 & 0.455 & 0.269 & 0.144 \\
& & (100.0\%) & (61.7\%) & (18.8\%) & (1.2\%)\\[5pt]
UNC &>,100.0 & -0.043 & -0.006 & 0.231 & 0.370 \\
& & (2.1\%) & (1.9\%) & (1.2\%) & (0.6\%)\\
& >,5.0 & 0.225 & 0.257 & 0.311 & 0.357 \\
& & (23.9\%) & (21.1\%) & (11.3\%) & (2.3\%)\\
& >,1.0 & 0.310 & 0.322 & 0.280 & 0.374 \\
& & (36.1\%) & (30.5\%) & (14.2\%) & (2.7\%)\\
& >,0.10 & 0.441 & 0.470 & 0.366 & 0.370 \\
& & (77.5\%) & (61.0\%) & (18.5\%) & (2.7\%)\\
& >,0.01 & 0.531 & 0.500 & 0.381 & 0.368 \\
& & (99.7\%) & (68.0\%) & (19.0\%) & (2.8\%)\\
&>,0.0 & 0.531 & 0.499 & 0.382 & 0.368 \\
& & (100.0\%) & (68.1\%) & (19.0\%) & (2.8\%)\\
\hline
\end{tabular*}
\end{table}

We next examined the size-duration correlation for only those TCP
connections used for HTTP protocols
(i.e., for connections more likely to be associated with a~user
involved in web browsing).
Table \ref{tab:corr2} gives the log--log correlation coefficients for
size and duration for HTTP connections
along with the percentage of connections contained in each
subpopulation. The results show slightly larger
but still weak positive correlations that are very comparable to those
for all TCP connections.

Overall, these results confirm the conclusions from prior studies
[\citet{r13}, \citet{r23}] that there is only weak correlation
between size and duration for TCP connections, even when considering
only those typically used for web browsing.
Furthermore, our analysis showed that these conclusions are essentially
independent of both size and duration thresholds.

\begin{figure*}
\centering
\begin{tabular}{c}

\includegraphics{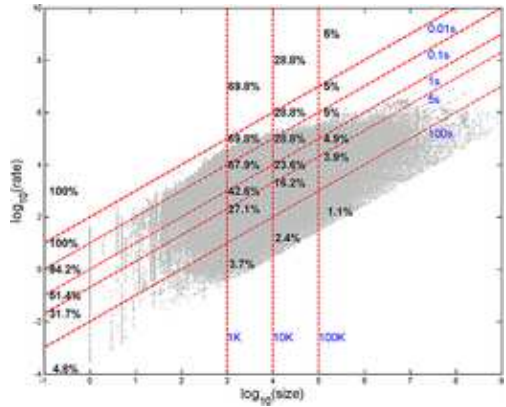}
\\
\footnotesize{(a) All Abilene connections, $n=1\mbox{,}318\mbox{,}661$.}\\[3pt]

\includegraphics{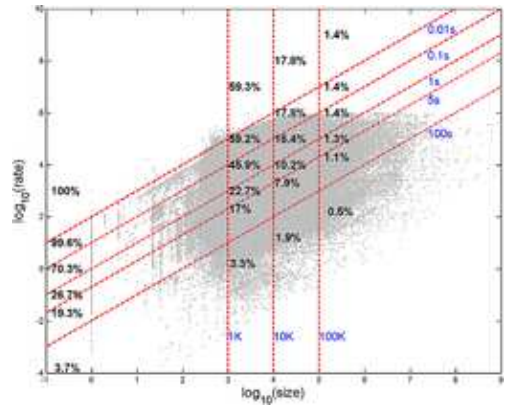}
\\
\footnotesize{(b) All Bell Labs connections, $n=2\mbox{,}313\mbox{,}744$.}\\[3pt]

\includegraphics{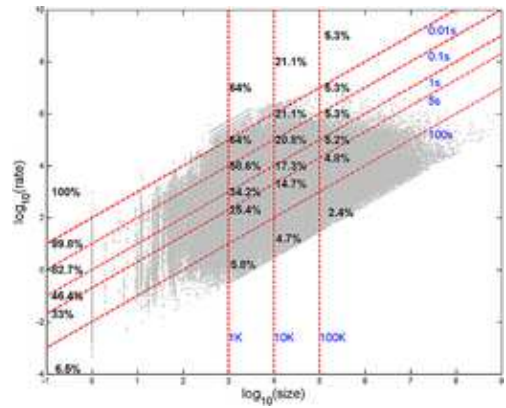}
\\
\footnotesize{(c) All UNC connections, $n=1\mbox{,}433\mbox{,}924$.}
\end{tabular}
\caption{Rate--size log--log correlation.}\label{fig:sr}
\end{figure*}

Size--rate relationships for all TCP connections are shown in Figure \ref
{fig:sr}. The horizontal axis is
$\log_{10}$(size) as before, but the vertical axis is $\log_{10}$(rate)
computed as
%
\begin{equation}
\log_{10}(\mathit{rate}) = \log_{10}(\mathit{size}/\mathit{duration}) = \log_{10}(\mathit{size}) - \log
_{10}(\mathit{duration}).
\label{eq:rate}
\end{equation}

The vertical lines represent the same size thresholds as before, but
the duration threshold lines now slope
upward to the right of the plot. Because of the minus sign in equation
(\ref{eq:rate}), the order of the threshold lines is reversed.
The intersections of the size and rate threshold lines with each other
or with the horizontal and vertical axes define
subpopulations that include all the points below (longer duration) and
to the right (larger size) of the intersection point
(note that the intersection of the vertical threshold lines with the
top horizontal axis defines a subpopulation based on size).
Using this different orientation, we see in Figure \ref{fig:sr}(a)
(Abilene) that 51.4\% of the connections had durations of more than
1 second and 28.8\% had a size of more than 10 kB. The subpopulation of
connections with durations greater than
1 second and size greater than 10 kB was 23.6\% of the total population.

\begin{table}
\caption{Log--log correlation coefficients of size and rate for all TCP
connections}\label{tab:corr3}
\begin{tabular*}{\textwidth}{@{\extracolsep{\fill}}lt{1.5}cccc@{}}
\hline
& \multicolumn{1}{c}{\textbf{Duration}\hspace*{3.9pt}} & \textbf{Size} & \textbf{Size} & \textbf{Size} & \textbf{Size} \\
& \multicolumn{1}{c}{\textbf{(seconds)}\hspace*{3.9pt}} & \multicolumn{1}{c}{$\bolds >$\textbf{0 kB}}
& \multicolumn{1}{c}{$\bolds >$\textbf{1 kB}}
& \multicolumn{1}{c}{$\bolds >$\textbf{10 kB}} & \multicolumn{1}{c@{}}{$\bolds >$\textbf{100 kB}}\\
\hline
Abilene & >,0.0 & 0.490 & 0.332 & 0.340 & 0.484 \\
& >,0.01 & 0.490 & 0.332 & 0.340 & 0.484 \\
& >,0.10 & 0.529 & 0.369 & 0.341 & 0.484 \\
& >,1.0 & 0.768 & 0.663 & 0.530 & 0.514 \\
& >,5.0 & 0.883 & 0.829 & 0.747 & 0.676 \\
& >,100.0 & 0.969 & 0.949& 0.928 & 0.902 \\[5pt]
Bell Labs & >,0.0 & 0.211 & 0.083 & 0.077 & 0.263 \\
& >,0.01 & 0.215 & 0.085 & 0.077 & 0.263 \\
& >,0.10 & 0.369 & 0.253 & 0.148 & 0.263 \\
& >,1.0 & 0.614 & 0.542 & 0.377 & 0.362 \\
& >,5.0 & 0.683 & 0.600 & 0.444 & 0.424 \\
& >,100.0 & 0.901 & 0.879 & 0.802 & 0.672 \\[5pt]
UNC & >,0.0 & 0.319 & 0.114 & 0.163 & 0.311 \\
& >,0.01 & 0.323 & 0.116 & 0.164 & 0.311 \\
& >,0.10 & 0.459 & 0.195 & 0.200 & 0.313 \\
& >,1.0 & 0.761 & 0.603 & 0.464 & 0.348 \\
& >,5.0 & 0.863 & 0.752 & 0.615 & 0.474 \\
& >,100.0 & 0.959 & 0.941 & 0.911 & 0.832 \\
\hline
\end{tabular*}
\end{table}

The size--rate correlation coefficients for each of the subpopulations
shown in Figure \ref{fig:sr} are given in Table \ref{tab:corr3}.
Considering all connections, we found only weak positive correlations
between size and rate across the three traces
(0.49, 0.211 and 0.319). When we used a 5-second duration threshold,
however, we found generally strong positive correlation
with coefficients of 0.883, 0.683 and 0.863. When the duration
threshold was increased to 100 seconds, the correlation
appeared to be even stronger (0.969, 0.901 and 0.959). Even more
interesting was the observation that for connections
with durations over 5 seconds, the correlation became weaker as this
subpopulation was further segmented
by larger values of the size thresholds. For example, in the UNC trace,
the size--rate correlation for connections
lasting longer than 5 seconds was reduced from 0.863 to 0.615 by
considering only those transferring
more than 10 kB and further reduced to 0.474 for a 100 kB threshold.
Further, if we applied a threshold only on size of 100 kB,
the correlation became even weaker in all traces (0.484, 0.263 and
0.311), leading to a completely different conclusion
about the relationship between size and rate than in \citet{r23}.

\begin{table}
\caption{Log--log correlation coefficients of size and rate for all HTTP
connections}
\label{tab:corr4}
\begin{tabular*}{\textwidth}{@{\extracolsep{\fill}}lt{1.5}cccc@{}}
\hline
& \multicolumn{1}{c}{\textbf{Duration}\hspace*{4pt}} & \textbf{Size} & \textbf{Size} & \textbf{Size} & \textbf{Size} \\
& \multicolumn{1}{c}{\textbf{(seconds)}\hspace*{4pt}} & \multicolumn{1}{c}{$\bolds >$\textbf{0 kB}}
& \multicolumn{1}{c}{$\bolds >$\textbf{1 kB}}
& \multicolumn{1}{c}{$\bolds >$\textbf{10 kB}}
& \multicolumn{1}{c@{}}{$\bolds >$\textbf{100 kB}}\\
\hline
Abilene & >,0.0 & 0.351 & 0.230 & 0.214 & 0.376 \\
& >,0.01 & 0.351 & 0.230 & 0.214 & 0.376 \\
& >,0.10 & 0.397 & 0.275 & 0.215 & 0.376 \\
& >,1.0 & 0.733 & 0.644 & 0.430 & 0.404 \\
& >,5.0 & 0.874 & 0.827 & 0.703 & 0.585 \\
& >,100.0 & 0.968 & 0.936 & 0.902 & 0.844 \\[5pt]
Bell Labs & >,0.0 & 0.093 & 0.057 & 0.040 & 0.224 \\
& >,0.01 & 0.097 & 0.058 & 0.040 & 0.224 \\
& >,0.10 & 0.274 & 0.222 & 0.111 & 0.224 \\
& >,1.0 & 0.606 & 0.554 & 0.339 & 0.337 \\
& >,5.0 & 0.698 & 0.636 & 0.418 & 0.417 \\
& >,100.0 & 0.920 & 0.901 & 0.805 & 0.668 \\[5pt]
UNC & >,0.0 & 0.176 & 0.142 & 0.156 & 0.286 \\
& >,0.01 & 0.179 & 0.144 & 0.158 & 0.286 \\
& >,0.10 & 0.341 & 0.230 & 0.204 & 0.291 \\
& >,1.0 & 0.718 & 0.657 & 0.508 & 0.339 \\
& >,5.0 & 0.837 & 0.798 & 0.664 & 0.506 \\
& >,100.0 & 0.962 & 0.952 & 0.922 & 0.860 \\
\hline
\end{tabular*}
\end{table}

Table \ref{tab:corr4} gives the log--log correlation coefficients for
size and rate for only those connections
likely to have been used for web access (HTTP protocols).
From examining the correlation coefficients for the entire population
and the subpopulations defined by different thresholds
on HTTP connections, we found essentially the same results as when
considering all connections---duration thresholds
lead to strong positive correlations between size and rate, and size
thresholds lead to weak correlations,
especially for the largest sizes. Thus, our results hold even for
connections used in web access
where user decisions might have a larger influence.

\begin{figure}

\includegraphics{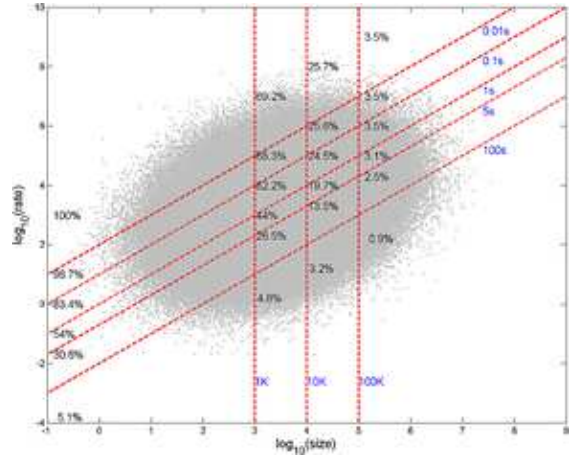}

\caption{Rate--size log--log correlation for simulated data.}\label{fig:simu}
\end{figure}

To investigate the performance of the thresholds for log--log
correlation in a more general context,
we considered a simulated bivariate Gaussian data set with
characteristics similar to the $\log_{10}$(size) and
$\log_{10}$(duration) pairs found in all connections from the UNC
trace. In particular, the same sample size was used,
and the mean vector and covariance matrix were estimated from the pairs
($\log_{10}$(size), $\log_{10}$(duration)).
The randomly generated values for sizes and durations were used in
equation (\ref{eq:rate})
to produce a scatter plot of the ($\log_{10}$(size), $\log_{10}$(rate))
bivariate distribution as shown in Figure \ref{fig:simu}
with the corresponding correlation coefficients for each subpopulation
given in Table \ref{tab:corr5}.

\begin{table}[b]
\caption{Log--log correlation coefficients for simulated size and rate data}
\label{tab:corr5}
\begin{tabular*}{\textwidth}{@{\extracolsep{\fill}}lt{1.5}cccc@{}}
\hline
& \multicolumn{1}{c}{\textbf{Duration}\hspace*{4pt}} & \textbf{Size} & \textbf{Size} & \textbf{Size} & \textbf{Size} \\
& \multicolumn{1}{c}{\textbf{(seconds)}\hspace*{4pt}} & \multicolumn{1}{c}{$\bolds >$\textbf{0 kB}}
& \multicolumn{1}{c}{$\bolds >$\textbf{1 kB}}
& \multicolumn{1}{c}{$\bolds >$\textbf{10 kB}}
& \multicolumn{1}{c@{}}{$\bolds >$\textbf{100 kB}}\\
\hline
Simulated & >,0.0 & 0.319 & 0.228 & 0.162 & 0.123 \\
& >,0.01 & 0.383 & 0.256 & 0.170 & 0.124 \\
& >,0.10 & 0.506 & 0.351 & 0.218 & 0.141 \\
& >,1.0 & 0.657 & 0.525 & 0.343 & 0.203 \\
& >,5.0 & 0.748 & 0.655 & 0.473 & 0.293 \\
& >,100.0 & 0.859 & 0.826 & 0.713 & 0.516 \\
\hline
\end{tabular*}
\end{table}

To illustrate that thresholds used for correlations are generally
unstable, the same analysis used above was applied to the simulated
data. The duration thresholds of 5 seconds, as used by \citet{r23},
resulted in a correlation of 0.748. This is slightly smaller than
the range of (0.83--0.88) they reported. The size thresholds of
100 kB resulted in the far smaller correlation of 0.123.

This shows that the dramatic differences in correlation, caused by the
different types of thresholds, are not data set specific.
In particular, it shows that the threshold effect we found on the
empirical data is not caused
by their non-Gaussian distributions of sizes and durations. If they
were, it would be difficult to generalize the relationships among
sizes, durations and rates obtained from our analysis due to different
thresholding effects. Even for this simulated log-normal data,
this same effect would be expected.

In fact, this finding is not surprising and can be analytically shown.
Correlation in a singly truncated bivariate
normal distribution was considered in \citet{r1}, and moments and
parameter estimation of a truncated
bivariate normal distribution was studied in \citet{r20}. Also,
correlation under linear constraints on the bivariate
range of the data in a multivariate normal distribution can be found in
\citet{r11}.
Let $(X,Y)$ be random variables whose joint density $f(x,y)$ is the
bivariate normal with means $(\mu_1,\mu_2)$,
variances $(\sigma_1^2,\sigma_2^2)$, and correlation coefficient $\rho
$. Then, the truncated (at $x>a$)
bivariate normal density is given by $f(x,y)/C$, where $C=P(X>a)$. If
$t=(a-\mu_1)/\sigma_1$, then
%
\[
\operatorname{Corr}(X,Y)=\rho\frac{\sqrt{1+ {te^{-t^2/2}}/{(\sqrt{2\pi} C)}
{-e^{-t^2}/{(2\pi C^2)} }}}
{\sqrt{1+ {\rho^2 te^{-t^2/2}}/{(\sqrt{2\pi} C)} -{\rho^2e^{-t^2}}/{(2\pi C^2)} }}
\]
for the truncated bivariate normal at $x>a$.

In summary, the type of thresholds used critically impacts the
correlation as expected.
The important issue of which threshold was shown in Section \ref
{sec:data} to be a choice of threshold on size.

\section{Correlation analysis of application data units}
\label{sec:corrd}

Our conclusion from the results reported in Section \ref{sec:corrt} is
that there is no strong (log--log) correlation of size and rate
or size and duration either in all TCP connections or only in those
associated with web browsing (HTTP protocols).
It may be the case, however, that considering the duration and size of
an entire TCP connection is too coarse a granularity
to see the effects of user behavior. There is a fundamental distinction
between the amount of data transferred
over a TCP connection and the size of a data object. Many application
protocols, notably HTTP,
reuse an already established TCP connection to transfer many data
objects of various sizes (a persistent connection).
Each of the data objects transferred might reflect a user's choice
while the total duration and number of objects and
bytes transferred on a persistent connection is more typically the
result of constraints on resources such as
memory at the server or browser program. Further, the duration of a
persistent connection may be dominated by application-level
synchronizations that can have a significant effect (for example, a web
browser waiting for the server's response,
human ``think times'' between requests, or application processing times
between sending data objects).
We can reduce the impact of all these effects by considering only the
size and duration of the actual transmissions
of individual data objects. The purpose of the analysis reported in
this section is to examine the possibility
that correlations between size and rate may exist for individual data
objects where they do not exist for entire TCP connections.

To understand the potential differences in correlation between rates
and sizes of connections vs. rates and sizes of data objects,
we extracted the sizes of individual data objects transferred from our
traces and repeated the analysis.
Methods have been developed to process packet-header traces and
identify the packets belonging to individual objects
(e.g., files, email messages) exchanged within a TCP connection.
Specifically, we used extensions of the methods
first described by \citet{r21} to identify the packets for all data
objects in the three traces used in this study.
Since these data objects are specific to the applications that use
them, we refer to them as application data units (ADUs).

For each ADU we defined its duration as the difference between the
timestamps of the first and the last packets seen
for this ADU in the trace. The size of an ADU was the total number of
bytes in the packets that belong to it.
Thus, size and duration of an ADU characterize the individual data
objects independent of how many were exchanged in a TCP connection.
As we did in Section \ref{sec:corrt}, we considered both the population
of all ADUs transferred on TCP connections
and the subpopulation of only those ADUs used in web browsing HTTP
protocols. In reporting the results, we omit
the details of size-duration correlations and concentrate on the
size--rate results, but the complete analysis is provided
at \url{http://www-dirt.cs.unc.edu/NetDepend/}. For the size-duration
correlations (both all ADUs and those in HTTP protocols)
we found generally the same results as for full connections---there
was only weak positive correlation,
typically in the range 0.10 to 0.50.

\begin{table}
\caption{Log--log correlation coefficients of size and rate for all ADUs}
\label{tab:corr6}
\begin{tabular*}{\textwidth}{@{\extracolsep{\fill}}lt{1.5}d{3.4}d{3.4}d{3.4}d{3.4}@{}}
\hline
& \multicolumn{1}{c}{\textbf{Duration}\hspace*{3.9pt}} & \multicolumn{1}{c}{\textbf{Size}}
& \multicolumn{1}{c}{\textbf{Size}} & \multicolumn{1}{c}{\textbf{Size}} & \multicolumn{1}{c@{}}{\textbf{Size}} \\
& \multicolumn{1}{c}{\textbf{(seconds)}\hspace*{3.9pt}} & \multicolumn{1}{c}{$\bolds >$\textbf{0 kB}}
& \multicolumn{1}{c}{$\bolds >$\textbf{1 kB}}
& \multicolumn{1}{c}{$\bolds >$\textbf{10 kB}} & \multicolumn{1}{c@{}}{$\bolds >$\textbf{100 kB}}\\
\hline
Abilene & >,0.0 & 0.679 & 0.095 & 0.046 & 0.180 \\
& & (100.0\%) & (60.3\%) & (15.9\%) & (2.7\%)\\
& >,0.01 & 0.768 & 0.328 & 0.151 & 0.180 \\
& & (82.6\%) & (45.4\%) & (14.9\%) & (2.7\%)\\
& >,0.10 & 0.828 & 0.533 & 0.309 & 0.180 \\
& & (59.3\%) & (29.1\%) & (12.8\%) & (2.7\%)\\
& >,1.0 & 0.907 & 0.841 & 0.702 & 0.348 \\
& & (18.1\%) & (7.7\%) & (5.1\%) & (2.2\%)\\
& >,5.0 & 0.963 & 0.938 & 0.876 & 0.684 \\
& & (6.6\%) & (3.9\%) & (2.4\%) & (1.1\%)\\
& >,100.0 & 0.993 & 0.973 & 0.954 & 0.921 \\
& & (0.4\%) & (0.2\%) & (0.2\%) & (0.1\%)\\[5pt]
Bell Labs & >,0.0 & 0.662 & 0.110 & 0.158 & 0.257 \\
& & (100.0\%) & (74.2\%) & (20.9\%) & (3.7\%)\\
& >,0.01 & 0.804 & 0.514 & 0.205 & 0.257 \\
& & (63.7\%) & (42.1\%) & (20.3\%) & (3.7\%)\\
& >,0.10 & 0.873 & 0.739 & 0.493 & 0.258 \\
& & (36.4\%) & (20.9\%) & (12.9\%) & (3.7\%)\\
& >,1.0 & 0.885 & 0.795 & 0.548 & 0.477 \\
& & (8.6\%) & (4.0\%) & (2.2\%) & (1.0\%)\\
& >,5.0 & 0.927 & 0.881 & 0.695 & 0.699 \\
& & (4.3\%) & (1.7\%) & (0.9\%) & (0.6\%)\\
& >,100.0 & 0.967 & 0.919 & 0.886 & 0.764 \\
& & (0.3\%) & (0.2\%) & (0.1\%) & (0.1\%)\\[5pt]
UNC & >,0.0 & 0.626 & 0.041 & -0.234 & 0.109 \\
& & (100.0\%) & (52.4\%) & (19.1\%) & (3.3\%)\\
& >,0.01 & 0.693 & 0.154 & -0.180 & 0.109 \\
& & (89.9\%) & (46.0\%) & (18.0\%) & (3.3\%)\\
& >,0.10 & 0.729 & 0.287 & -0.039 & 0.122 \\
& & (72.3\%) & (33.1\%) & (14.8\%) & (3.3\%)\\
& >,1.0 & 0.838 & 0.647 & 0.412 & 0.252 \\
& & (34.3\%) & (14.8\%) & (8.4\%) & (3.1\%)\\
& >,5.0 & 0.909 & 0.802 & 0.640 & 0.455 \\
& & (16.4\%) & (9.9\%) & (6.2\%) & (2.7\%)\\
& >,100.0 & 0.988 & 0.940 & 0.928 & 0.846 \\
& & (2.6\%) & (2.0\%) & (1.9\%) & (1.3\%)\\
\hline
\end{tabular*}
\end{table}

Table \ref{tab:corr6} shows the log--log correlation coefficients for
all ADUs in the three traces
along with the percentage of connections contained in each
subpopulation.
When considering all ADUs, we found moderately stronger positive
size--rate correlation with coefficients 0.679, 0.662 and 0.626,
consistent with the fact that we eliminated additional factors that
affect the duration of TCP connections.
When a~duration threshold of 5 seconds was applied, the positive
correlation of size and rate became very strong
(0.963, 0.927 and 0.909). When we applied instead a~size threshold of
100 kB, we found only very weak correlations
(0.18, 0.257 and 0.109).\looseness=1

\begin{figure*}
\centering
\begin{tabular}{c}

\includegraphics{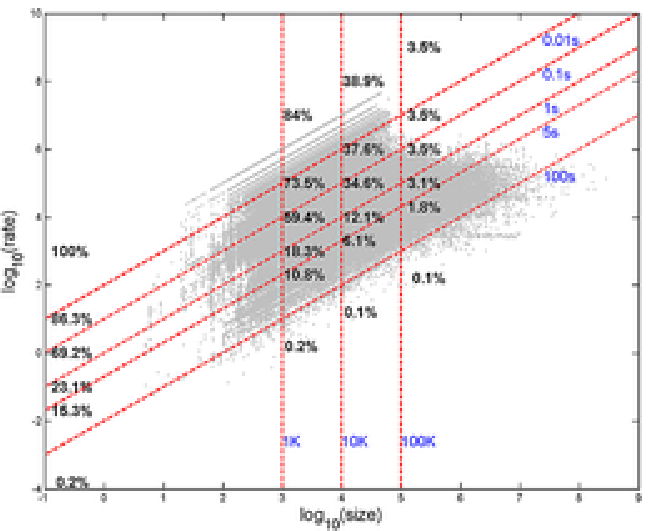}
\\
\footnotesize{(a) Abilene data units (HTTP), $n=4\mbox{,}464\mbox{,}446$.}\\[3pt]

\includegraphics{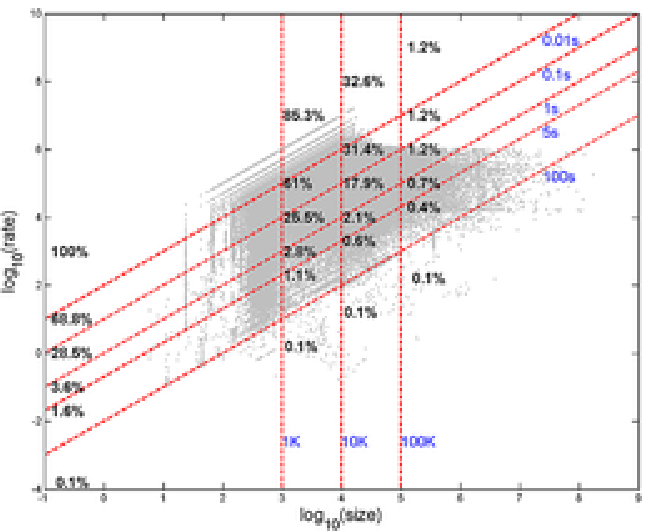}
\\
\footnotesize{(b) Bell Labs data units (HTTP), $n=7\mbox{,}462\mbox{,}332$.}\\[3pt]

\includegraphics{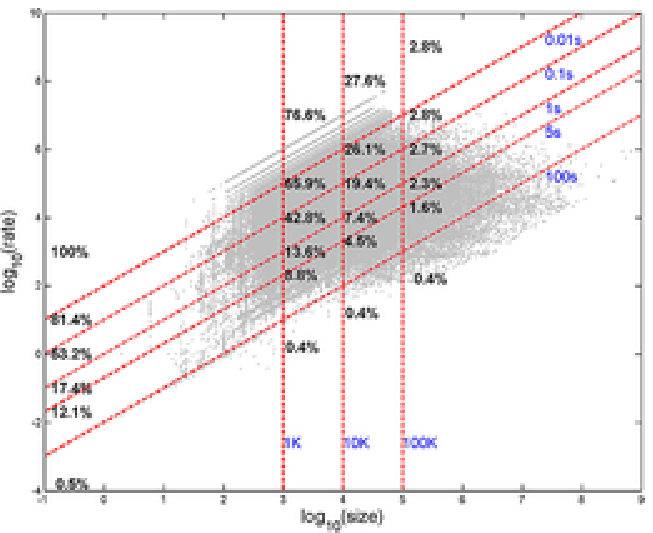}
\\
\footnotesize{(c) UNC data units (HTTP), $n=4\mbox{,}331\mbox{,}676$.}
\end{tabular}
\caption{Rate--size log--log correlation.}\label{fig:srdata}
\end{figure*}

Figure \ref{fig:srdata} shows the scatter plots for those ADUs in TCP
connections used for HTTP protocols
in the three traces of $\log_{10}$(rate) vs. $\log_{10}$(size)
with the usual threshold lines. The corresponding log--log correlation
coefficients
are given in Table \ref{tab:corr7}. The log--log correlation coefficients
for size and rate over all HTTP data units were 0.356, 0.249 and 0.346,
showing much weaker positive correlation than for all ADUs.
When a duration threshold of 5 seconds was applied, the positive
correlation became very strong (0.979, 0.946 and 0.926).
If we applied instead a size threshold of 100 kB, we found correlations
ranging from very weakly negative ($-0.132$ and $-0.003$)
to very weakly positive (0.124).

Overall, these results show that there is little correlation between
size and rate when considering object sizes
(especially those larger than about 100 kB) and not just total TCP
connection sizes. This is a new result and, we believe,
is a stronger result because it reflects more directly the properties
of data objects which users might choose to transfer
(or not transfer) based on their expected transfer rates.

\section{Extremal dependence analysis}
\label{sec:eda}

It has been shown in a number of places [see \citet{r9} and references
therein for example] that distributions for
Internet flows are heavy tailed. This fact casts doubt on the use of Pearson's
correlation for understanding correlation among the largest values,
since it only represents an average of values close to the mean, which
can be insensitive
to a very few relatively large values. In this section we consider an
alternative dependence measure
that focuses on the largest values of the bivariate random variables
(size/rate, size/duration) with heavy tailed distribution.
It is based on the idea that the extremes may carry distinctive
dependence information, which is not seen from moderate values
for bivariate heavy tailed data. Extremal dependence analysis assesses
the tendency of large values of components of a random vector
with heavy tailed distribution to occur simultaneously.
The information from extremal dependence analysis can be qualitatively
different from Pearson's correlation.
In addition, understanding the extremes is particularly interesting
since we intuitively expect users to be more conscious of
network performance when transferring the largest object. If we show
the absence of such a feature in our data sets, it would indicate that no
such effect is significant when considering a broad population of users
and network conditions.

There has been a growing interest on measuring tail dependence between
two random variables;
see \citet{r8}, Ledford and Tawn (\citeyear{r14,r15}), \citet{r19}, \citet{r22} and the references therein.
These methods are capable of providing a more stable and robust
approach than the usual Pearson's correlation.
Recently \citet{r22} considered the quotient correlation as an
alternative measure to Pearson's correlation.
It can be viewed as the correlation coefficient in extreme value theory
and provides more intuitive
information where the tail behavior of data matters. It
shows more efficiency over other linear correlation measures when
nonlinear dependence occurs.
A study of the tail behavior and nonlinear relationships among size,
duration and rate using the quotient correlation
would be interesting and we intend to do a thorough analysis in the future.

\begin{table}
\caption{Log--log correlation coefficients of size and rate for HTTP ADUs}
\label{tab:corr7}
\begin{tabular*}{\textwidth}{@{\extracolsep{\fill}}lt{1.5}cd{2.3}d{2.3}d{2.3}@{}}
\hline
& \multicolumn{1}{c}{\textbf{Duration}\hspace*{3.9pt}} & \multicolumn{1}{c}{\textbf{Size}} & \multicolumn{1}{c}{\textbf{Size}}
& \multicolumn{1}{c}{\textbf{Size}} & \multicolumn{1}{c@{}}{\textbf{Size}} \\
& \multicolumn{1}{c}{\textbf{(seconds)}\hspace*{3.9pt}} & \multicolumn{1}{c}{$\bolds >$\textbf{0 kB}}
& \multicolumn{1}{c}{$\bolds >$\textbf{1 kB}}
& \multicolumn{1}{c}{$\bolds >$\textbf{10 kB}}
& \multicolumn{1}{c@{}}{$\bolds >$\textbf{100 kB}}\\
\hline
Abilene & >,0.0 & 0.356 & 0.081 & 0.011 & 0.124 \\
& >,0.01 & 0.576 & 0.318 & 0.088 & 0.124 \\
& >,0.10 & 0.679 & 0.461 & 0.165 & 0.125 \\
& >,1.0 & 0.921 & 0.843 & 0.663 & 0.273 \\
& >,5.0 & 0.979 & 0.958 & 0.907 & 0.675 \\
& >,100.0 & 0.986 & 0.977 & 0.932 & 0.836 \\[5pt]
Bell Labs & >,0.0 & 0.249 & -0.035 & -0.020 & -0.132 \\
& >,0.01 & 0.631 & 0.428 & 0.042 & -0.132 \\
& >,0.10 & 0.761 & 0.611 & 0.266 & -0.130 \\
& >,1.0 & 0.887 & 0.799 & 0.553 & 0.251 \\
& >,5.0 & 0.946 & 0.914 & 0.740 & 0.578 \\
& >,100.0 & 0.931 & 0.904 & 0.873 & 0.803 \\[5pt]
UNC & >,0.0 & 0.346 & 0.115 & -0.139 & -0.003 \\
& >,0.01 & 0.534 & 0.290 & -0.069 & -0.003 \\
& >,0.10 & 0.649 & 0.442 & 0.105 & 0.016 \\
& >,1.0 & 0.882 & 0.792 & 0.565 & 0.169 \\
& >,5.0 & 0.926 & 0.875 & 0.721 & 0.456 \\
& >,100.0 & 0.968 & 0.917 & 0.909 & 0.860 \\
\hline
\end{tabular*}
\end{table}

Usual methods in the context of extreme value theory require the
distribution of the coordinatewise sample maxima
under certain centering and scaling to converge to a product measure.
This makes it hard to find meaningful
results about the tail behavior of the product of two random variables
due to their broad definitions.
We use a new advance in statistical methodology proposed by \citet{r8}, \citet{r19}
in this paper because it is based only on assuming multivariate regular
variation
of the observation vector and is a distribution moment instead of the
measure of a region in the positive quadrant.

In the following subsections we briefly introduce our approach and give
results from applying it to the data sets used in our study.
Our primary purpose is to determine if a similar conclusion about the
relationship between size and rate
described in the previous two sections can be obtained using a
completely different statistical methodology
based on extreme value theory.

\subsection{Background}


In extreme value theory the concept of extremal (or asymptotic)
independence is designed to
make the asymptotic, limiting distribution of extremes a product distribution.
Extremal independence implies that for a bivariate random vector with a
heavy tailed distribution,
the probability of both variables being large simultaneously is
negligible in comparison to the probability of one of them being large.
In other words, extreme values of the two variables tend to occur
separately, not simultaneously.
See \citet{r2} for a formal introduction of extremal independence,
and \citet{r18} for an overview of recent work in this area.

\begin{figure*}
\centering
\begin{tabular}{c}

\includegraphics{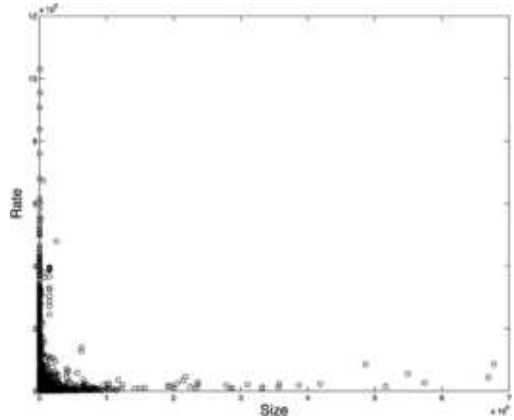}
\\
\footnotesize{(a) Scatter plot.}\\[3pt]

\includegraphics{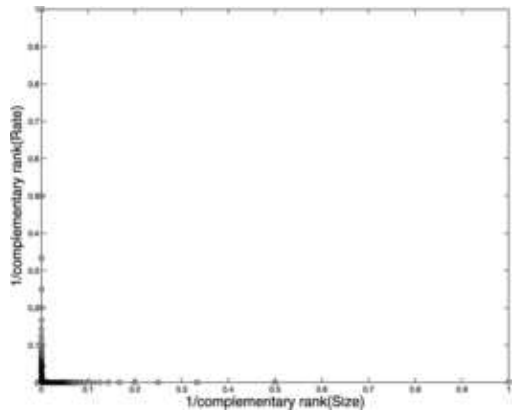}
\\
\footnotesize{(b) ICRT transform.}\\[3pt]

\includegraphics{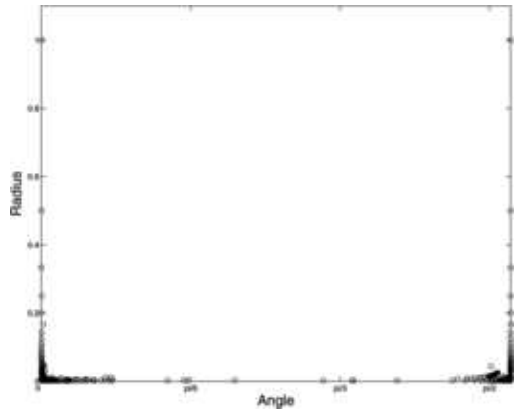}
\\
\footnotesize{(c) Polar coordinates plot with a 5\% threshold.}
\end{tabular}
\caption{Rate vs. size of UNC connection trace.}\label{fig:icrt}
\end{figure*}

In Figure \ref{fig:icrt}(a), rate versus size of the UNC connection
trace is displayed as an example of extremal independence.
There are some connections with a very large size (rate), but not
unusually large rates (sizes, respectively).
Thus, the large values of size and rate do not tend to occur together,
which implies the extremal
independent case. This phenomenon was called ``axis hugging'' in \citet
{r8} since the data tend to hug the axes,
and there is a very large empty region in the upper right corner of the
plot. This is in line with the results
provided in the previous two sections.
This phenomenon can be found in other areas. For example, in finance
an important issue is whether large changes in exchange rate returns
for different currencies tend to occur together or separately;
see \citet{r3}, \citet{r17}, \citet{r19}. Environmental statistics, including the study
of extrema of sea and wind conditions,
is another area where extremal (in)dependence analysis is likely to be
useful; see \citet{r4}, Ledford and Tawn (\citeyear{r14,r15}).

From here on we explain our extremal dependence analysis procedure step
by step.
One practical problem in applying the concept of extremal dependence is
that the two variables should be on
similar scales or transformed to the standard case where the marginal
distribution tails of two
variables are asymptotically equivalent and regularly varying with
index $-1$; see \citet{r8}.
As seen in Figure \ref{fig:icrt}(a), scaling is not comparable in this
example because the variables are of different orders
of magnitude. Furthermore, there are many large values in some
directions, and fewer in others.
To make axes comparable, \citet{r8} suggested two ways for the
estimation in the nonstandard case,
the Inverse Complementary Rank Transform (ICRT) and
angular rank methods. The angular rank method needs extra effort of
tail index normalization, which brings much uncertainty.
Therefore, we use the ICRT, which is based on the complementary ranks
for both marginal distributions simultaneously,
but in a way that preserves the critical bivariate structure [\citet{r4}, \citet{r5}, \citet{r10}].
Figure \ref{fig:icrt}(b) shows the scatter plot of the transformed
data from (a). The ICRT was applied to both marginal distributions
of size and rate.

After the ICRT transformation the data are represented in polar coordinates
and then thresholded to the subset with largest radius components.
Thresholds are needed because we are interested in the behavior of the
more extreme values of two variables.
Extreme values are defined as the large radius components of polar
coordinates and the thresholds determine ``how large'' they should be.
For example, assuming that there are 1 million data points, a 0.01\%
threshold selects the subset of the 100 largest radii,
while a 10\% threshold selects the largest 100,000 radii. Then the
distribution of the angles corresponding to the
exceedances is studied for indications of extremal independence.
Figure \ref{fig:icrt}(c) shows the data in polar coordinates with a
5\% threshold.
An important consequence of the polar coordinate representation is
simple quantification of the idea of ``axis hugging.''
In particular, distributions that have extremal independence are
characterized by most of the angles $\theta$ (horizontal axis)
being close to the endpoints of the interval $[0, \pi/2]$ as observed
in Figure \ref{fig:icrt}(c).
On the other hand, distributions that have large values
occurring simultaneously will have a different distribution of angles,
in particular, with a greater frequency of angles
near the middle of the range $[0, \pi/2]$.

The Extremal Dependence Measure (EDM), introduced in \citet{r8},
quantifies the concepts of extremal dependence and axis hugging,
by measuring dependency between large values of two variables. EDM is
defined through a set of angles
$\theta_1, \theta_2,\ldots,\theta_k \in[0,\pi/2]$, which are obtained
from polar transformations of the data:
\[
\operatorname{EDM}=1-\biggl(\frac{4}{\pi}\biggr)^2 \frac{1}{k} \sum_{i=1}^k
\biggl(\theta_i-\frac{\pi}{4}\biggr)^2.
\]
Here the parameter $k$ is the number of observations whose modulus is
greater than some threshold value.

Using the selected set, we then compute the EDM,
which measures the dependence between two variables as described above.
The basis of the EDM is the mean squared distance
from the data angles to $\pi/4$, the center of the range of possible
values, but it is linearly adjusted so that
its values correspond to familiar values for the usual correlation. In
particular, when the data points hug the axes
(essentially extremal independence), most of the angles are near 0 or
$\pi/2$, so $\frac{1}{k} \sum_{i=1}^k
(\theta_i-\pi/4)^2 \approx(\frac{\pi}{4})^2$,
and $\operatorname{EDM} \approx0$.
When the data points lie near the 45 degree line,
$\frac{1}{k} \sum_{i=1}^k (\theta_i-\pi/4)^2 \approx0$ and
$\operatorname{EDM} \approx1$ (extremal dependence).
One more indicator for interpretation of EDM comes from the fact that
when the data have angles that are nearly uniformly
distributed on $[0, \pi/2]$, a simple calculation shows that $\frac
{1}{k} \sum_{i=1}^k (\theta_i-\pi/4)^2 \approx
\frac{1}{3}(\frac{\pi}{4})^2$ and $\operatorname{EDM} \approx2/3$.
We use the uniform
distribution to interpret EDM, because it is in between the cases of
extremal dependence and independence.

\subsection{Data analysis}

We applied EDA to the data sets used for this study and obtain EDM
values but present here only the results for the size--rate analysis.
In addition to EDM, a careful study of the full distribution of angles
would provide useful information.
The complete analysis for all the data sets can be found at
\url{http://www-dirt.cs.unc.edu/NetDepend/}.
We used a range of percentages of the data sets,
selecting the largest 0.01\%, 0.02\%, 0.05\%, 0.1\%, 0.2\%, 0.5\%, 1\%,
2\%, 5\%, 10\% and 20\% subpopulations of connections
when their joint size and rate values are transformed to polar
coordinate space. Figure \ref{fig:EDM} shows the EDM values
as a function of the subpopulation size expressed as the percentages
given above. Each plot shows results for four cases:
all connections, only HTTP connections, all ADUs, and only ADUs from
HTTP. With only one exception, the largest 10\%
of connections in each case show a tendency toward extremal
independence between size and rate since $\mathrm{EDM}<0.4$,
which is smaller than the uniform reference value of $\mathrm{EDM}=2/3$. This
suggests less piling
at the ends than for the uniform.\looseness=1

The exception is for all connections in the Abilene trace where the
results might be considered inconclusive.
Even the largest 20\% of connections have EDM values indicating
extremal independence (a few cases might be considered inconclusive).
In no case is there an EDM value that would indicate any strong
extremal dependence between size and rate ($\mathrm{EDM}>0.75$),
even in the largest 20\% of joint size and rate values. The results for
size and duration were similar except
with a few more values falling into the inconclusive range.

\section{Conclusion}
\label{sec:summary}

Our major results are as follows:
\begin{enumerate}[]
\item[$\bullet$] We found that correlation between size and duration is much
weaker than one might expect.
\item[$\bullet$] In contrast to \citet{r23}, we did not find strong correlation
between size and rate in TCP transfers.
This result holds for the total size of data transferred in all TCP
connection, for the total size only in web connections and for the
sizes of individual data objects (e.g., file sizes).
\begin{figure*}
\centering
\begin{tabular}{c}

\includegraphics{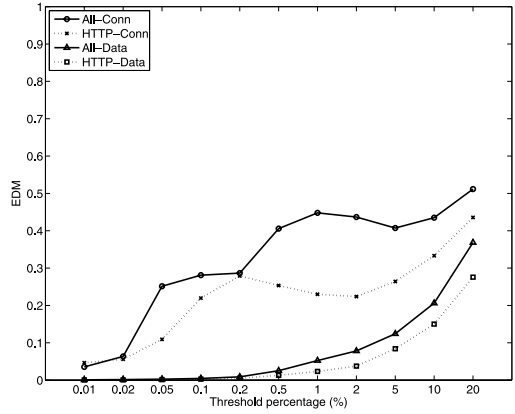}
\\
\footnotesize{(a) Abilene.}\\[3pt]

\includegraphics{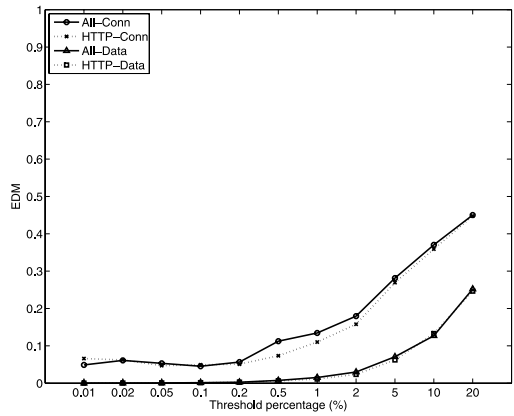}
\\
\footnotesize{(b) Bell Labs.}\\[3pt]

\includegraphics{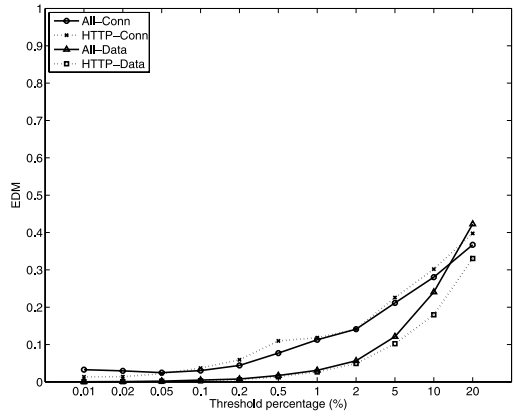}
\\
\footnotesize{(c) UNC.}
\end{tabular}
\caption{EDM of rate--size correlation for Abilene, Bell Labs and UNC traces.}\label{fig:EDM}
\end{figure*}
\item[$\bullet$] We explained our differing conclusion by examining the impact of
thresholding variables
on the log--log correlation coefficient. We showed that the finding
of correlation in \citet{r23} was primarily due to the use of a
duration threshold, which artificially created a strong correlation.
Based on TCP connection startup mechanisms, we argue that thresholds
on size should be more useful than thresholds on duration in the
analysis of correlations.
\item[$\bullet$] Our extremal dependence analysis also showed a tendency toward
extremal independence between size and rate.
\end{enumerate}

An important observation from the results presented here is that
there is no reason to believe that size and rate are strongly
correlated in TCP connections. This conclusion was strengthened by
showing that the same lack of strong correlation between size and
rate exists even when considering the entire size of data
transferred in all TCP connections, the sizes only in HTTP
connections or the sizes of individual data objects (e.g., file
sizes). It was also strengthened by studying the dependence with
two very different methods of statistical analysis. As a consequence,
our results indicated that available transfer rates had no measurable
effect on object selection. In other words, we did not find that
users are less willing to download larger objects as their available
bandwidth decreases. Our observation is about the statistics for the
entire population, which is the important part of traffic modeling,
not about whether some users are responsive to available bandwidth.
We are not claiming that users never make such choices. We have all
made such choices ourselves. We claim only that we found no
conclusive evidence that this was a common element of user behavior
in networked applications. The near independence of size and rate
may have important implications for network operators about
technology deployment decisions, especially at the network edge.
Further, we are not claiming that our findings about correlation of
size and rate in TCP connections hold for the Internet of today.
The traces examined in this study are quite old (circa 2002)
relative to the pace of change in Internet technology and
applications since then. While examining this issue with recent traces is
the subject of future work, TCP's slow start and congestion avoidance
remain unchanged and we have no reason
to expect any fundamental change in user behavior in response to
network bandwidth.

Our primary contributions in this paper concern best practices in the
use of statistical
methods for studying Internet flows and we demonstrated that care in
applying statistical methodology is
essential. We saw that different interpretations of the data are
possible depending on the choice of thresholds. We recommend that
when thresholds are used to segment data sets, the analysis should
consider the joint effects of the thresholds in a manner similar to
the two-dimensional analysis of duration and size used here. We
also recommend the use of EDA as a method to examine the
relationships among the subset of larger values of a joint
distribution (e.g., size and rate).

\section*{Acknowledgments}

The authors would like to thank the Editor and two referees for
suggestions that helped
very much to improve this article.

\printaddresses


\begin{thebibliography}{99}
\bibitem[\protect\citeauthoryear{Aitkin}{1964}]{r1}
\textsc{Aitkin, M.} (1964).
Correlattion in a singly truncated bivariate normal distribution.
\textit{Psychometrika} \textbf{29} 263--270.
\MR{0170412}

\bibitem[\protect\citeauthoryear{Beirlant et~al.}{2004}]{r2}
\textsc{Beirlant, J., Goegebeur, Y., Segers, J.} and \textsc{Teugels,
J.} (2004).
\textit{Statistics of Extremes: Theory and Applications}.
Wiley, Chichester.
\MR{2108013}

\bibitem[\protect\citeauthoryear{Coles, Heffernan and Tawn}{1999}]{r3}
\textsc{Coles, S., Heffernan, J.} and \textsc{Tawn, J.} (1999).
Dependence measures for extreme value analyses.
\textit{Extremes} \textbf{2} 339--365.

\bibitem[\protect\citeauthoryear{de~Haan and de~Ronde}{1998}]{r4}
\textsc{de~Haan, L.} and \textsc{de~Ronde, J.} (1998).
Sea and wind: Multivariate extremes at work.
\textit{Extremes} \textbf{1} 7--45.
\MR{1652944}

\bibitem[\protect\citeauthoryear{Einmahl, de~Haan and Piterbarg}{2001}]{r5}
\textsc{Einmahl, J., de~Haan, L.} and \textsc{Piterbarg, V.} (2001).
Nonparametric estimation of the spectral measure of an extreme value
distribution.
\textit{Ann. Statist.} \textbf{29} 1401--1423.
\MR{1873336}

\bibitem[\protect\citeauthoryear{Hern\'{a}ndez-Campos}{2006}]{r7}
\textsc{Hern\'{a}ndez-Campos, F.} (2006).
Generation and validation of empirically-derived tcp application
workloads.
Unpublished Ph.D. dissertation, Univ. North Carolina at
Chapel Hill, Dept. Computer Science.

\bibitem[\protect\citeauthoryear{Hern\'{a}ndez-Campos et~al.}{2004}]{r9}
\textsc{Hern\'{a}ndez-Campos, F., Marron, J.~S., Samorodnitsky, G.} and
\textsc{Smith,
F.~D.} (2004).
Variable heavy tails in internet traffic.
\textit{Journal of Performance Evaluation} \textbf{58} 261--284.

\bibitem[\protect\citeauthoryear{Hern\'{a}ndez-Campos et~al.}{2005}]{r8}
\textsc{Hern\'{a}ndez-Campos, F., Jeffay, K., Park, C., Marron, J.~S.}
and \textsc{Resnick,
S.} (2005).
Extremal dependence: Internet traffic applications.
\textit{Stoch. Models} \textbf{21} 1--35.
\MR{2124357}

\bibitem[\protect\citeauthoryear{Huang}{1992}]{r10}
\textsc{Huang, X.} (1992).
Statistics of bivariate extreme values.
Ph.D. thesis, Tinbergen Institute Research Series 22, Erasmus
Univ. Rotterdam, Postbus 1735, 3000DR, Rotterdam, The Netherlands.

\bibitem[\protect\citeauthoryear{Johnson and Kotz}{1972}]{r11}
\textsc{Johnson, N.~L.} and \textsc{Kotz, S.} (1972).
\textit{Distributions in Statistics: Continuous Multivariate
Distributions}.
Wiley, New York.
\MR{0418337}

\bibitem[\protect\citeauthoryear{Kurose and Ross}{2007}]{r12}
\textsc{Kurose, J.~F.} and \textsc{Ross, K.~W.} (2007).
\textit{Computer Networking: A Top-Down Approach}.
Addison Wesley, Boston, MA.

\bibitem[\protect\citeauthoryear{Lan and Heidemann}{2006}]{r13}
\textsc{Lan, K.-C.} and \textsc{Heidemann, J.} (2006).
A measurement study of correlations of internet flow characteristics.
\textit{Computer Networks} \textbf{50} 46--62.

\bibitem[\protect\citeauthoryear{Ledford and Tawn}{1996}]{r14}
\textsc{Ledford, A.~W.} and \textsc{Tawn, J.~A.} (1996).
Statistics for near independence in multivariate extreme values.
\textit{Biometrika} \textbf{83} 169--187.
\MR{1399163}

\bibitem[\protect\citeauthoryear{Ledford and Tawn}{1997}]{r15}
\textsc{Ledford, A.~W.} and \textsc{Tawn, J.~A.} (1997).
Modelling dependence within joint tail regions.
\textit{J. Roy. Statist. Soc. Ser. B} \textbf{59}
475--499.
\MR{1440592}

\bibitem[\protect\citeauthoryear{Lu et~al.}{2005}]{r16}
\textsc{Lu, D., Qiao, Y., Dinda, P.~A.} and \textsc{Bustamante, F.~E.} (2005).
Characterizing and predicting tcp throughput on the wide area
network.
In \textit{Proceedings of IEEE International Conference on Distributed
Computing Systems 2005}, 414--424.

\bibitem[\protect\citeauthoryear{Poon, Rockinger and Tawn}{2001}]{r17}
\textsc{Poon, S.-H., Rockinger, M.} and \textsc{Tawn, J.} (2001).
New extreme-value dependence measures and finance applications.
In \textit{CEPR Discussion Paper No. 2762}.
Available at Social Science Research Network:
\url{http://ssrn.com/abstarct=267283}.

\bibitem[\protect\citeauthoryear{Resnick}{2002}]{r18}
\textsc{Resnick, S.} (2002).
Hidden regular variation, second order regular variation and
asymptotic variation.
\textit{Extremes} \textbf{5} 303--336.
\MR{2002121}

\bibitem[\protect\citeauthoryear{Resnick}{2004}]{r19}
\textsc{Resnick, S.} (2004).
On the foundations of multivariate heavy tailed analysis.
\textit{J. Appl. Probab.} \textbf{41A} 191--212.
\MR{2057574}

\bibitem[\protect\citeauthoryear{Rosenbaum}{1960}]{r20}
\textsc{Rosenbaum, S.} (1960).
Moments of a truncated bivariate normal distribuion.
\textit{J. Roy. Statist. Soc. Ser. B} \textbf{23}
405--408.
\MR{0132630}

\bibitem[\protect\citeauthoryear{Smith et~al.}{2001}]{r21}
\textsc{Smith, F.~D., Hern\'{a}ndez-Campos, F., Jeffay, K.} and \textsc
{Ott, D.} (2001).
What tcp/ip protocol headers can tell us about the web.
In \textit{Proceedings of ACM SIGMETRICS 2001 Conference}
245--256.

\bibitem[\protect\citeauthoryear{Zhang, Breslau and Shenker}{2002}]{r23}
\textsc{Zhang, Y., Breslau, L., V., P.} and \textsc{Shenker, S.} (2002).
On the characteristics and origins of internet flow rates.
In \textit{Proceedings of ACM SIGCOMM 2002 Conference}
309--322.\

\bibitem[\protect\citeauthoryear{Zhang}{2008}]{r22}
\textsc{Zhang, Z.} (2008).
Quotient correlation: A sample based alternative to pearson's
correlation.
\textit{Ann. Statist.} \textbf{36} 1007--1030.
\MR{2396823}

\end{thebibliography}
\end{document}